\newcommand{\bee}{\begin{equation}}
\newcommand{\ene}{\end{equation}}
\newcommand{\beea}{\begin{eqnarray}}
\newcommand{\enea}{\end{eqnarray}}

 \documentclass[aps,prl,showpacs,superscriptaddress]{revtex4-1}
 \usepackage{graphicx}
 \usepackage{dcolumn}
 \usepackage{bm}
 
\begin{document}
 \title{Theory for large amplitude electrostatic ion shocks in quantum plasmas}
 \author{M. Akbari-Moghanjoughi}
 \affiliation{Azarbajan University of Shahid-Madani, Faculty of Sciences, Department of Phyics, 51745-406 Tabriz, Iran}
\author{P. K. Shukla}
\address{International Centre for Advanced Studies in Physical Sciences \& Institute for Theoretical Physics, Ruhr University Bochum, D-44780 Bochum, Germany}
\address{Department of Mechanical and Aeospace Engineering \& Center for Energy Research, University of
California San Diego, La Jolla, CA 92093, U. S. A.}

\begin{abstract}
We present a generalized nonlinear theory for large amplitude electrostatic (ES) ion shocks in collisional quantum plasmas composed of mildly coupled degenerate electron fluids of arbitrary degeneracy and non-degenerate strongly correlated ion fluids with arbitrary atomic number. For our purposes, we use the inertialess electron momentum equation including the electrostatic, pressure gradient and relevant quantum forces, as well as a generalized viscoelastic momentum (GVEM) equation for strongly correlated non-degenerate ions. The ion continuity equation, in the quasi-neutral approximation, then closes our nonlinear system of equations. When the electric field is eliminated from the GVEM equation by using the inertialess electron momentum equation, we then obtain a generalized GVEM and  the ion  continuity equations exhibiting nonlinear couplings between the ion number density and the ion fluid velocity. The pair of nonlinear equations is numerically solved to study the dynamics of arbitrary large amplitudes planar and non-planar ES shocks arising from the balance between harmonic generation nonlinearities and  the ion fluid viscosity for a wide range of the plasma mass-density and the ion atomic-number that are relevant for the cores of giant planets (viz. the Jupiter) and compact stars (viz. white dwarfs). Our numerical results reveal that the ES shock density profiles strongly depend on the plasma number density and composition (the atomic-number) parameters. Furthermore, density  perturbations propagate with Mach numbers which significantly dependent on the studied plasma fractional parameters. It is concluded that the dynamics of the ES shocks in the super-dense degenerate plasma is quite different in the core of a white dwarf star from that in the lower density crust region.
\end{abstract}
\pacs{52.27.Gr,52.30.Ex, 52.35.-g, 52.35.Fp, 52.35.Tc}

\maketitle

\section{Introduction}

Recently, there has been a great deal of interest \cite{Manfredi2005, Shukla2006, Brodin2007,Brodin2008,Shukla2010, Shukla2011, Vladimirov2011, Haas2011} in studying various collective phenomena in dense quantum plasmas which are ubiquitous in a wide-range of physical systems, including compact astrophysical objects (e.g. the cores of white dwarfs \cite{Chandrasekhar1931, Chandrasekhar1935} and neutron stars \cite{chai}), giant planetary cores (e.g. the interior of the planet Jupiter \cite{Fortov2000},  high-energy density compressed plasmas created by powerful laser beams \cite{Glenzer2007}, as well as semiconductors \cite{Mark,Gardner} and metallic nanostructures \cite{PRB2008} for high-technological applications, table-top quantum free-electron lasers (FEL) \cite{FEL} for producing coherent tunable x-rays and gamma rays, and quantum diodes \cite{Ang}, etc. Quantum plasmas are usually composed of degenerate electron and nondegenerate ion fluids. Since, in quantum plasmas the number density of degenerated electrons, $n_0$, is rather high, the average inter-electron/ion distances $d \simeq(3/4\pi n_0)^{1/3}$ can be comparable with the electron thermal de Broglie wavelength, $2 \pi \hbar/\sqrt{m_e V_T}$, which is much smaller than the Landau length $L  = e^2/k_B T_e$. Here where $e$ is the magnitude of the electron charge, $k_B$ the Boltzmann constant, $T_e$ the thermal electron temperature, $\hbar$ the Planck constant divided by $2\pi$, $m_e$ the rest mass of the electron, and $V_T = (k_B T_e/m_e)^{1/2}$ the electron thermal speed. In quantum plasmas, the wavefunctions of degenerated fermionic ingredients (electrons) overlap giving rise to the Fermi-Dirac distribution function and density of states (DoS). In fact, the importance of the quantum electron degeneracy was recognized eighty years ago by Fowler \cite{Fowler} and Chandrasekhar \cite{Chandrasekhar1931} in the study of non-relativistic and ultra-relativistic pressures for degenerate electrons that were required for setting up the hydrostatic stability of a gravitating star that eventually becomes a compact star called a white dwarf. Clearly, in quantum plasmas, the equations of state (EoS) significantly differ from classical plasmas where non-degenerate electrons obey the ordinary Maxwell-Boltzmann distribution function. In quantum plasmas, the electron and ion gamma factors (which are defined as the ratio of the Coulomb potential to the thermal energy of the plasma particles) are given as $\Gamma_e =e^2/d K_B T_e \equiv L/d$ and $\Gamma_i =Z^2e^2/d k_B T_i$, respectively, being smaller and much larger than unity, implying a weak correlation for degenerate electrons and strong correlations for non-degenerate ion fluids in the quasineutral plasma regime.

The collective behavior of a quantum Fermi plasmas with a statistical ensemble of degenerate electrons has been investigated in the pioneering papers of Klimontovich and Silin \cite{Klimontovich}, and later by Bohm and Pines \cite{Bohm}, who have used the quantum kinetic Wigner and Poisson equations to obtain the dispersion relation for quantum electron-plasma oscillations (EPOs). The frequency spectra of the latter
can be shown to be, $\omega =(\omega_{pe}^2 + k^2V_{Fe}^2 +\hbar^2k^4/4 m_e)^{1/2}$, where $\omega_{pe} = (4\pi n_0 e^2/m_e)^{1/2}$ is the electron plasma frequency and $V_{Fe} =(\hbar/m_e)(3\pi^2 n_0)^{1/3}$ is the Fermi electron speed. The frequency spectra of quantum EPOs have also been derived \cite{Manfredi2001, Manfredi2005} from the quantum electron fluid equations \cite{Zphysik}, composed of the electron continuity, non-relativistic electron momentum equation including the quantum statistical pressure and the quantum recoil effect, and Poisson's equation. On the other hand, in his classic experimental work, Glenzer {\it et al.} \cite{Glenzer2007} have conclusively demonstrated the quantum mechanical collective effect by measuring the scattering of electromagnetic waves off the quantum EPOs in
a high-energy density (warm dense matter) laboratory plasma. Numerical analyses of Poisson's equation and the nonlinear Schr\"odinger equation for non-relativistic degenerate electron fluids revealed the formation of a dark soliton and two-dimensional quantized vortices \cite{Shukla2006}, as well as nano-structures \cite{Dastgeer2007} in quantum electron plasmas with immobile ions.

On the other hand, inclusion of the ion dynamics leads to electrostatic oscillations (ESOs) supported by the restoring force due to the quantum statistical pressure, the quantum Bohm force and the inertia of non-degenerate ions. By using Poisson's equation, the momentum equation for non-relativistic degenerate inertialess electrons and the classical hydrodynamic equations (the continuity and momentum equations) for non-correlated ions, Hass {\it et al} \cite{Haas2003} derived a modified Korteweg-de Vries (KdV) equation
for small amplitude nonlinear ion oscillations in an unmagnetized quantum plasma. The KdV equation admits a localized solitary pulse as a stationary solution. Large amplitude localized ion oscillations in  a collisionless quantum plasma were investigated by many authors \cite{ShuklaEliassonJPP2008, EliassonShuklaEPL2012, Akbari2010} ignoring the ion correlation effects. The latter and the ion fluid viscosity have recently been incorporated in the small amplitude shock wave theory \cite{Ghosh2012} of ion oscillations in a quantum plasma.

In this paper, we present a theory for fully nonlinear nonstationary ES ion shock waves in a quantum viscous plasma with mildly coupled degenerate electrons and strongly correlated ion fluids for a wide-range of the Chandrasekhar's relativity parameter and the plasma atomic number. For our purposes, we use the hydrodynamic force balance (the electric force balancing the gradient of the degenerate electron pressure, gradient of the potentials involving the electron-exchange and electron-correlations, the gradient of the Bohm potential) for inertialess electrons and the generalized viscoelastic ion momentum (GVIM) equation, together with the ion continuity equation, invoking the quasi-neutrality approximation. The GVIM equation contains the essential physics of the ion correlation decay rate, as well as of the ion fluid bulk and shear viscosities. When the electric force is eliminated from the GVIM equation by using the electric field, from the momentum equation for inertialess degenerate electron fluids, we obtain a pair of equations exhibiting the coupling between the ion fluid velocity and the ion number density. The pair of equations are then numerically solved to study the dynamics of IS ion shocks in planar as well as non-planar geometries. Our numerical analysis exhibits distinguished features for the formation of ES shock structures in different ion mass-density ranges which might play a very important role in acceleration of charged particles in dense quantum plasmas, such as those in the cores of white dwarf stars and in compressed plasmas created by intense laser and electron beams.

\section{The viscoelastic hydrodynamics model}\label{Model}

We consider a fully-ionized quantum plasma composed of strongly correlated non-degenerate warm ion and mildly coupled degenerate electron fluids. Since, the ion thermal energy is much smaller than the electron Fermi-electron energy, we shall not consider quantum forces on the ion fluid. The viscoelastic hydrodynamic (VEHD) model for such an electron-ion plasma is governed by a set of equations \cite{Shukla2011}

\begin{equation}\label{vector}
\begin{array}{l}
\frac{{d{n_i}}}{{dt}} + {n_i}\nabla \cdot{{\bf{u}}_i} = 0,\hspace{3mm}\frac{d}{{dt}} = \frac{\partial }{{\partial t}} + {\bf{u}}\cdot\nabla , \\
\left( {1 + {\tau _\eta }\frac{d}{{dt}}} \right)\left[ {{m_i}{n_i}\frac{{d{{\bf{u}}_i}}}{{dt}} - Ze{n_i}{\bf{E}}
+ \nabla {P_i} + Z{n_i}{m_i}{\nu _{ie}}\delta {\bf{u}}} \right] \\
= \left[ {{\eta _v}{\nabla ^2}{{\bf{u}}_i} + \left( {\xi  + \frac{{{\eta _v}}}{3}} \right)\nabla (\nabla \cdot{{\bf{u}}_i})} \right],
\hspace{3mm}\delta {\bf{u}} = {{\bf{u}}_i} - {{\bf{u}}_e}\\0 =  - e{n_e}{\bf{E}} - \nabla {P_{tot}}
+ {m_e}{n_e}{\nu _{ei}}\delta {\bf{u}}, \\
\end{array}
\end{equation}

where  $\tau_\eta$, $\nu_{ei}$($\nu_{ie}$), $\eta_v$, and $Z$ are the viscoelastic relaxation time, electron-ion (ion-electron) collision frequency, shear viscosity, and the ion charge state, respectively. Also, the total electron pressure, $P_{tot}=P_e+P_C+P_{xc}$, consists of pressures due to the electron degeneracy, Coulomb interaction, and electron-exchange interaction, respectively. The electron viscosity has been shown to become dominant only at much larger densities, such as in neutron-star crusts \cite{Itoh}, hence, we will ignore it in the following analysis. It is also known that in quantum plasmas the electron-electron collisions are rare due to the Pauli blocking-mechanism \cite{Shukla2010}. Therefore, it is convincing to only consider the ion-fluid viscosity effects in our model.

We simplify the model by supposing that degenerate electrons are inertialess and the plasma is quasi-neutral, viz. $n_e\simeq Zn_i$. Thus, we have $\rho\simeq m_i n_i$, and obtain

\begin{equation}
\label{scalar}
\begin{array}{l}
\frac{{d\rho }}{{dt}} + \rho \nabla \cdot{\bf{u}} = 0, \\
\left( {1 + {\tau _\eta }\frac{d}{{dt}}} \right)\left[\rho{\frac{{d{\bf{u}}}}{{dt}} + \nabla {P_G}}\right]
= \left[ {{\eta _v}{\nabla ^2}{\bf{u}} + \left( {\xi  + \frac{{{\eta _v}}}{3}} \right)\nabla (\nabla \cdot{\bf{u}})} \right], \\
\end{array}
\end{equation}

where, $P_G=P_{tot}+P_i$ is the generalized plasma pressure. Moreover, it is convenient to express all involved parameters in terms of the Chandrasekhar's relativity parameter, $R=P_{Fe}/m_e c=(\rho/\rho_c)^{1/3}\simeq (\rho_6/ \mu_e)^{1/3}$, where $\rho_c\simeq 2\times 10^6$ gr/cm$^3$ and $\mu_e=A/Z\simeq 2$ are the normalizing plasma mass-density and number of nucleons per electron with $A$ and $Z$ being the atomic weight and number, and $P_{Fe}$ is the electron relativistic Fermi momentum.

Other quantities can be expressed in terms of the relativity parameter $R$ and the plasma atomic-number $Z$. For instance, for the electron degeneracy pressure, we have  \cite{Chandrasekhar1939}

\begin{equation}\label{p}
P_{e} = \frac{{\pi m_e^4{c^5}}}{{3{h^3}}}\left[ {R\left( {2{R^2} - 3} \right)\sqrt {1 + {R^2}}
+ 3{{\sinh}^{ - 1}}R} \right],
\end{equation}

while, for the Coulomb interaction pressure in the spherical Wigner-Seitz cell approximation, we have \cite{Salpeter}

\begin{equation}\label{ctf}
{P_C} =  - \frac{{8{\pi ^3}m_e^4{c^5}}}{{{h^3}}}\left[ {\frac{{\alpha {Z^{2/3}}}}{{10{\pi ^2}}}{{\left( {\frac{4}{{9\pi }}} \right)}^{1/3}}} \right]{R^4},\hspace{3mm}\alpha  = \frac{{{e^2}}}{{\hbar c}} \simeq \frac{1}{{137}}.
\end{equation}

Finally, for the electron exchange interaction \cite{Salpeter}, one may write

\begin{equation}
\label{ex}
\begin{array}{l}
{P_{xc}} =  - \frac{{2\alpha m_e^4{c^5}}}{{{h^3}}}\left\{ {\frac{1}{{32}}\left( {{\beta ^4} + {\beta ^{ - 4}}} \right) + \frac{1}{4}\left( {{\beta ^2} + {\beta ^{ - 2}}} \right) - \frac{3}{4}\left( {{\beta ^2} - {\beta ^{ - 2}}} \right)\ln \beta } \right. - \frac{9}{{16}} + \frac{3}{2}{\left( {\ln \beta } \right)^2} \\
\left. { - \frac{R}{3}\left( {1 + \frac{R}{{\sqrt {1 + {R^2}} }}} \right)\left[ {\frac{1}{8}\left( {{\beta ^3} - {\beta ^{ - 5}}} \right) - \frac{1}{4}\left( {\beta  - {\beta ^{ - 3}}} \right) - \frac{3}{2}\left( {\beta  + {\beta ^{ - 3}}} \right)\ln \beta  + \frac{{3\ln \beta }}{\beta }} \right]} \right\},
\end{array}
\end{equation}

where, $\beta = R + \sqrt{1 + R^2}$. On the other hand, the ion-viscosity is  \cite{Radha}

\begin{equation}
\label{eta}
\eta_v  \simeq \frac{{{\rho _c}{R^5}}}{{Z{I_2}(1 + {R^2})}},\hspace{3mm}{I_2} = \sqrt {\frac{\pi }{3}} \ln {Z^{1/3}} + \frac{2}{3}\ln \left( {1.3 + \frac{{2.3}}{{\sqrt \Gamma  }}} \right) - \frac{{1 + 2{R^2}}}{{2 + 2{R^2}}} + \frac{{0.27{R^2}}}{{1 + {R^2}}},
\end{equation}

\begin{figure}[ptb]\label{Figure1}
\includegraphics[scale=.6]{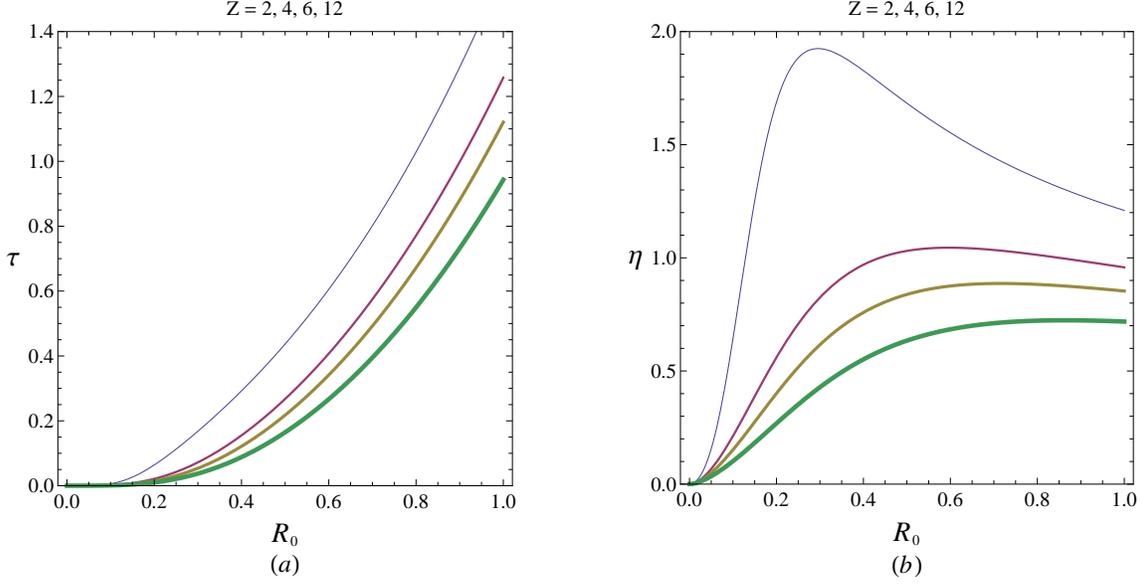}\caption{Figure 1 shows the variations in strongly-coupled viscosity parameters of relativistically degenerate plasma ion-fluid, such as the viscoelastic relaxation-time, $\tau$, and the plasma ion fluid-viscosity, $\eta$ with respect to different mass-density (relativistic-degeneracy parameter, $R_0$) and the atomic-number, $Z$. The curve thickness in plots indicate the increase in the varied parameter in each plot.}
\end{figure}

\noindent where, $\Gamma=(Ze)^2/(r_0 k_B T)$ is the coupling parameter with $r_0=(3/4\pi n_0)^{1/3}$ being the Wigner-Seitz spherical-cell radius containing only one electron on average. Variations of the parameters $I_1$ and $I_2$ are given in Ref. \cite{Radha} for various coupling parameter values. According to DeWitt \cite{DeWitt}, the strongly coupled plasma is in liquid form for the range of the plasma ion coupling of $\Gamma<178$ to which we focus our attention in the following analysis. The plasma coupling factor, in terms of the relativity parameter for a typical white dwarf temperature of $T\simeq10^6 K$ is approximately $\Gamma  = 22.75{Z^{5/3}}R$, valid for a wide ange of the plasma mass-density and atomic-number variations. The viscoelastic relaxation-time is directly related to the plasma viscosity via the following relation \cite{Radha}

\begin{equation}
\label{rel}
{\tau _\eta } = \frac{{5{m^*}\eta }}{{{n_e}P_{Fe}^2}},\hspace{3mm}{m^*} \simeq {m_e}\sqrt {1 + {R^2}},
\end{equation}

where, $m^*$ is the free electron effective-mass. The viscoelastic relaxation time is thus given as the following, with the fitting constants presented in Ref. \cite{Radha}

\begin{equation}
\label{tau}
{\tau _\eta } = \frac{{\pi {\hbar ^3}}}{{4ZI_2{e^4}{m_e}\sqrt {1 + {R^2}} }} \simeq \frac{2 \times {10^{ - 17}}}{{ZI_2\sqrt {1 + {R^2}} }}.
\end{equation}

Furthermore, ignoring the ion pressure compared to that of the electrons, we may write the plasma generalized pressure in terms of a generalized effective potential \cite{Akbari2012}

\begin{equation}
\label{dimensional}
\nabla \Psi_G  = \frac{1}{c_s^2\rho }\nabla P_G(R) = \frac{1}{c_s^2\rho }\frac{{dP_G(R)}}{{dR}}\nabla R,\hspace{3mm}\Psi_G  = \frac{1}{c_s^2}\int {\frac{{{d_R}P_G(R)}}{\rho }} dR,
\end{equation}

with the effective plasma potential, defined as

\begin{equation}
\label{psi}
{\Psi _G} = \sqrt {1 + {R^2}}  - \beta R + \frac{\alpha }{{2\pi }}\left( {R - \frac{{3{{\sinh }^{ - 1}}R}}{{\sqrt {1 + {R^2}} }}} \right),
\end{equation}
where, $\beta= {3^{1/3}}({{2\alpha }}/{5})({{{2Z}}/{\pi }})^{2/3}$. The dimensionless set of VEHD equations, Eqs. (\ref{scalar}), can be obtained by using the following scalings

\begin{equation}\label{normal}
\nabla \to \frac{{{c_{s}}}}{{{\omega _{pi}}}}\bar \nabla,\hspace{3mm}t \to \frac{{\bar t}}{{{\omega _{pi}}}},\hspace{3mm}\rho \to \bar \rho{\rho_0},\hspace{3mm}u \to \bar u{c_{s}},\hspace{3mm} R \to R_0\bar \rho^{1/3},
\end{equation}
where, the normalizing factors, $\rho_0$, ${\omega _{pi}} = (e/m_i)\sqrt {{4\pi}{\rho_{0}}}\simeq 10^{18}Z\sqrt{R_0^3}$ (with $R_0=(\rho_0/\rho_c)^{1/3}$ and $\rho_c\simeq 2\times 10^6 gr/cm^3$), and ${c_{s}} = c\sqrt {{m_e}/{m_i}}$, denote the equilibrium plasma mass-density, characteristic ion plasma frequency and the plasma sound speed, respectively. Thus, the normalized equations, dropping the bar notation for simplicity, read

\begin{equation}
\label{dimensionless}
\begin{array}{l}
{d_t}\rho  + \rho \nabla \cdot{\bf{u}} = 0, \\
(1 + \tau {d_t})\left[ {\rho \left( {{d_t}{\bf{u}} + \nabla {\Psi _G}} \right)} \right]
= \eta\left[ {{\nabla ^2}{\bf{u}} + \frac{1}{3}\nabla (\nabla \cdot{\bf{u}})} \right]. \\
\end{array}
\end{equation}

where, ${\tau}={\omega _{pi}}{\tau _\eta }\simeq 1.92\sqrt {R_0^3} /(I_2\sqrt {1 + {R_0^2}\rho^{2/3}})$ and $\eta  = ({\omega _{pi}}/c_s^2){\eta _v} \simeq 2.07R_0^{7/2} /[{I_2}{\rho ^{1/3}}(1 + R_0^2{\rho ^{2/3}})]$. It should be noted that, at very high densities, such as in the neutron star crust region, the electron-electron collisions can be very important dominating the electron viscosity over that of  the ions, the consideration of which is beyond the scope of the present investigation. Equations (\ref{dimensionless}) can be considered as the most general model for an unmagnetized viscoelastic fully interacting quantum plasma, valid for a wide- range of the plasma mass density, atomic-number, and the ion-fluid coupling factor. The quantum plasma parameters range of interest covering the planet interiors and white dwarfs are approximately, $0<R_0<1$, $1<Z<12$ and $1<\Gamma<178$. The quantum fluid model, following the closed equation-set relevant to different plasma geometries for the one-dimensional (planar ($\nu=0$), cylindrical ($\nu=1$), and spherical ($\nu=2$)) coordinate system, can be written as

\begin{equation}\label{simp}
\begin{array}{l}
\frac{{\partial \rho (r,t)}}{{\partial t}} + \frac{1}{{{r^\nu }}}\frac{\partial }{{\partial r}}\left[ {{r^\nu }\rho (r,t)u(r,t)} \right] = 0, \\
\left[ {1 + \tau (\rho ,Z)\frac{d}{{dt}}} \right]\left[ {\rho (r,t)\left( {\frac{{du(r,t)}}{{dt}} + \frac{{\partial {\Psi _G}(\rho ,Z)}}{{\partial r}}} \right)} \right] \\ = \eta (\rho ,Z)\left[ {\frac{1}{{{r^\nu}}}\frac{\partial }{{\partial r}}\left( {{r^\nu}\frac{{\partial u(r,t)}}{{\partial r}}} \right) + \frac{1}{3}\frac{\partial }{{\partial r}}\left( {\frac{1}{{{r^\nu}}}\frac{{\partial {r^\nu}u(r,t)}}{{\partial r}}} \right)} \right]. \\
\end{array}
\end{equation}

In the following sections, we will evaluate the characteristics of the viscoelastic ES nonlinear excitations
in terms of fractional Fermi-Dirac plasma parameters. Note that, in our model, we have ignored the bulk viscosity, $\zeta$, compared to that of shear, $\eta$, due to negligible contributions from compressibility
of high pressure plasma fluids.

Figure 1 displays the variations of (a) the normalized viscoelastic relaxation time, $\tau$, and (b) the normalized shear ion fluid-viscosity, $\eta$ with respect to the plasma degeneracy parameter, $R_0$, for different plasma atomic-number, $Z$. It is observed from Fig. 1(a) that for all atomic-number values the viscoelastic relaxation time increases with  the increase of the plasma number density, related to the relativistic degeneracy parameter $R_0$, and this increase is smaller for larger values of plasma atomic-number, $Z$. Figure 1(b), on the other hand, reveals that the plasma ion-viscosity first increases with the increase of the plasma number density up to a maximum value and then decreases again. This behavior is observed for all plasma atomic-number value being more pronounced for plasmas with lower atomic numbers. It is also remarked that the relative value of the ion fluid-viscosity parameter for a given plasma mass density is always lower for heavier element composed plasmas.

\section{Linear and nonlinear density perturbations}\label{Sagdeev}

\begin{figure}[ptb]\label{Figure2}
\includegraphics[scale=.6]{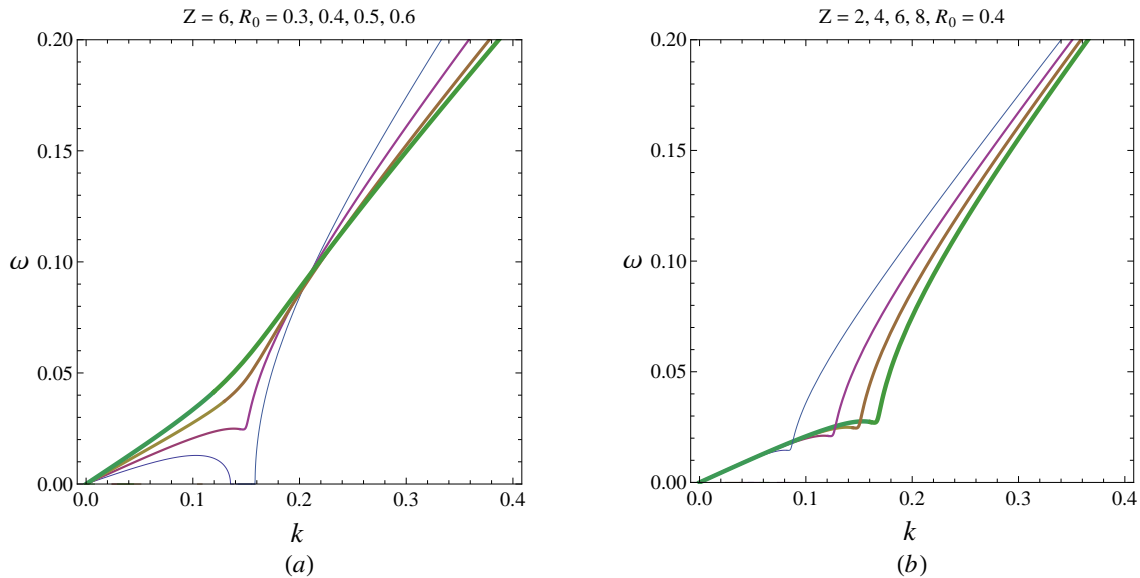}\caption{Figure 2 shows the linear wave dispersion of ion  waves in the relativistically degenerate strongly-coupled viscous plasma and the variation of wave speeds in terms of the ion  mass-density (relativistic-degeneracy parameter $R_0$) and the atomic-number $Z$. The curve thickness in plots indicates the increase in the varied parameter in each plot.}
\end{figure}

Important information can be obtained considering the plasma response to linear perturbations. This is accomplished by the Fourier analysis of Eq. (\ref{simp}), and by using the harmonic operators $\partial_r=ik$ and $\partial_t=-i\omega$. Accordingly, we have the following first order relations

\begin{equation}\label{rel}
\begin{array}{l}
k{u_1} - \omega {\rho _1} = 0, \\
3(i + \tau \omega )(k{\rho_1}T - {u_1}\omega ) + 4\eta {k^2}{u_1} = 0,
\end{array}
\end{equation}
where

\begin{equation}\label{t}
T = \frac{{R_0^2}}{{3\sqrt {1 + R_0^2} }} - \frac{{{R_0}\beta }}{3} + \frac{\alpha }{{2\pi }}\left( {\frac{{{R_0}}}{3} - \frac{{{R_0}}}{{1 + R_0^2}} + \frac{{R_0^2{{\sinh }^{ - 1}}{R_0}}}{{{{(1 + R_0^2)}^{3/2}}}}} \right).
\end{equation}

From Eqs. (\ref{rel}) we readily obtain the generalized dispersion relation

\begin{equation}\label{dr}
3 i {k^2}T + {k^2}(4\eta  + 3T\tau )\omega  - 3 i {\omega ^2} - 3\tau {\omega ^3} = 0.
\end{equation}

Figure 2 depicts the linear dispersion of viscoelastic ion waves in a relativistically degenerate plasma and the variation of the phase-speed for different plasma mass density and atomic-numbers. Figure 2(a) reveals that the linear wave dispersion is fundamentally governed by the plasma mass density. Below a critical plasma density, a dispersion gap opens in the wave-number range, indicated for the thin curve in the plot. As the plasma density increases above the critical value, for a fixed plasma composition, the dispersion gap disappears, but there still exists a sharp discontinuity in the linear wave speed. Also, as the plasma mass-density increases further this discontinuity disappears. Such anomalous feature may be attributed to the interplay of strong plasma interactions and ion fluid couplings expected for a super-dense degenerate plasma. Figure 2(b) shows the effect of the plasma composition on the linear wave dispersion. It is evident that in the small and long-wavelength limits, the linear wave speed is not affected by the change to plasma composition, while this is converse in the medium-wavelength region due to the presence of the mentioned discontinuity in the plasma wave speed.

\begin{figure}[ptb]\label{Figure3}
\includegraphics[scale=.6]{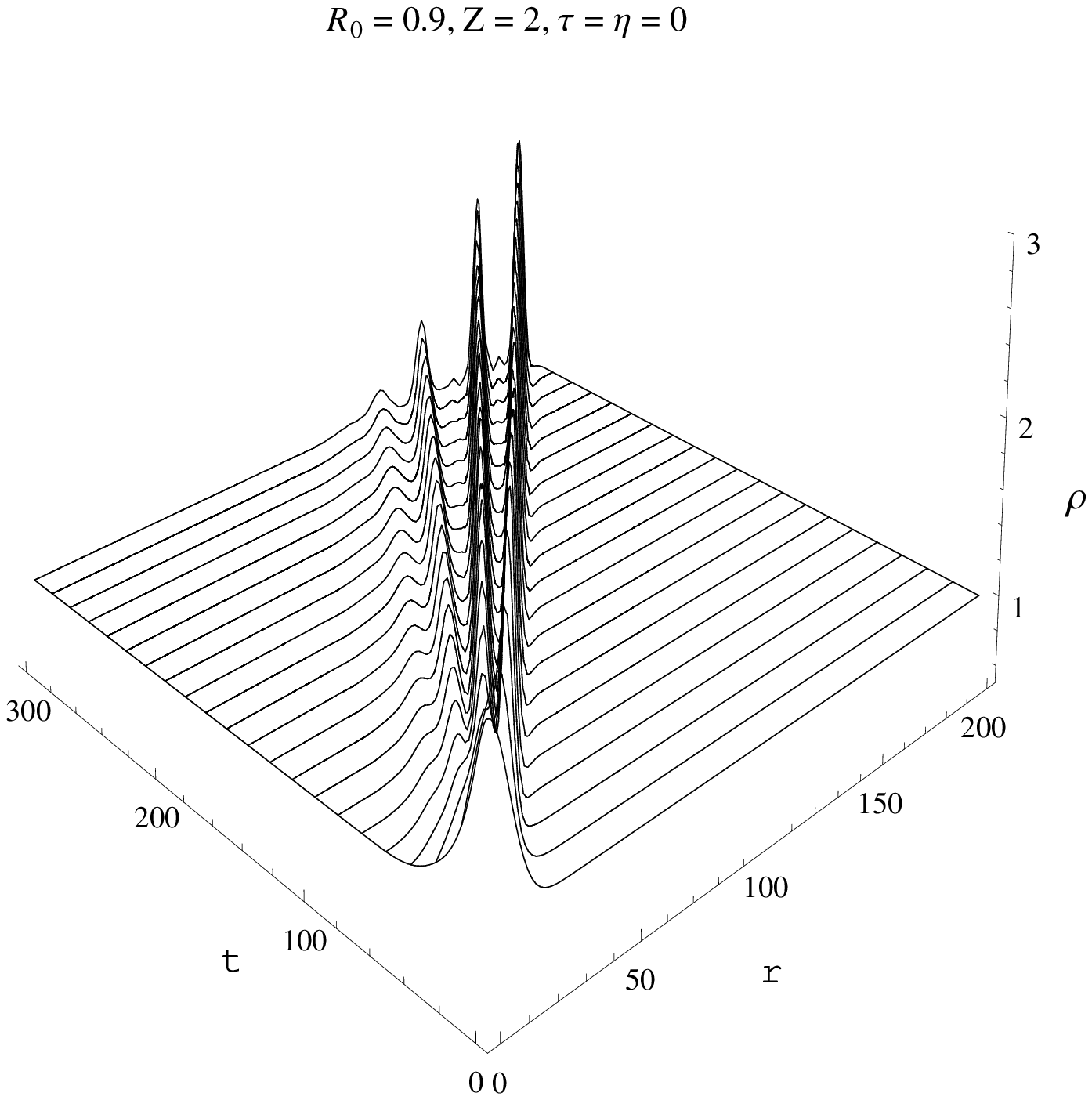}\caption{Figure 3 shows the evolution of a Gaussian density pulse in a non-viscous ($\eta=\tau=0$) collisional quantum plasma for a given plasma number density ($n_0\simeq 4.3\times 10^{29}cm^{-3}$) and atomic-number ($Z=2$).}
\end{figure}

\begin{figure}[ptb]\label{Figure4}
\includegraphics[scale=.6]{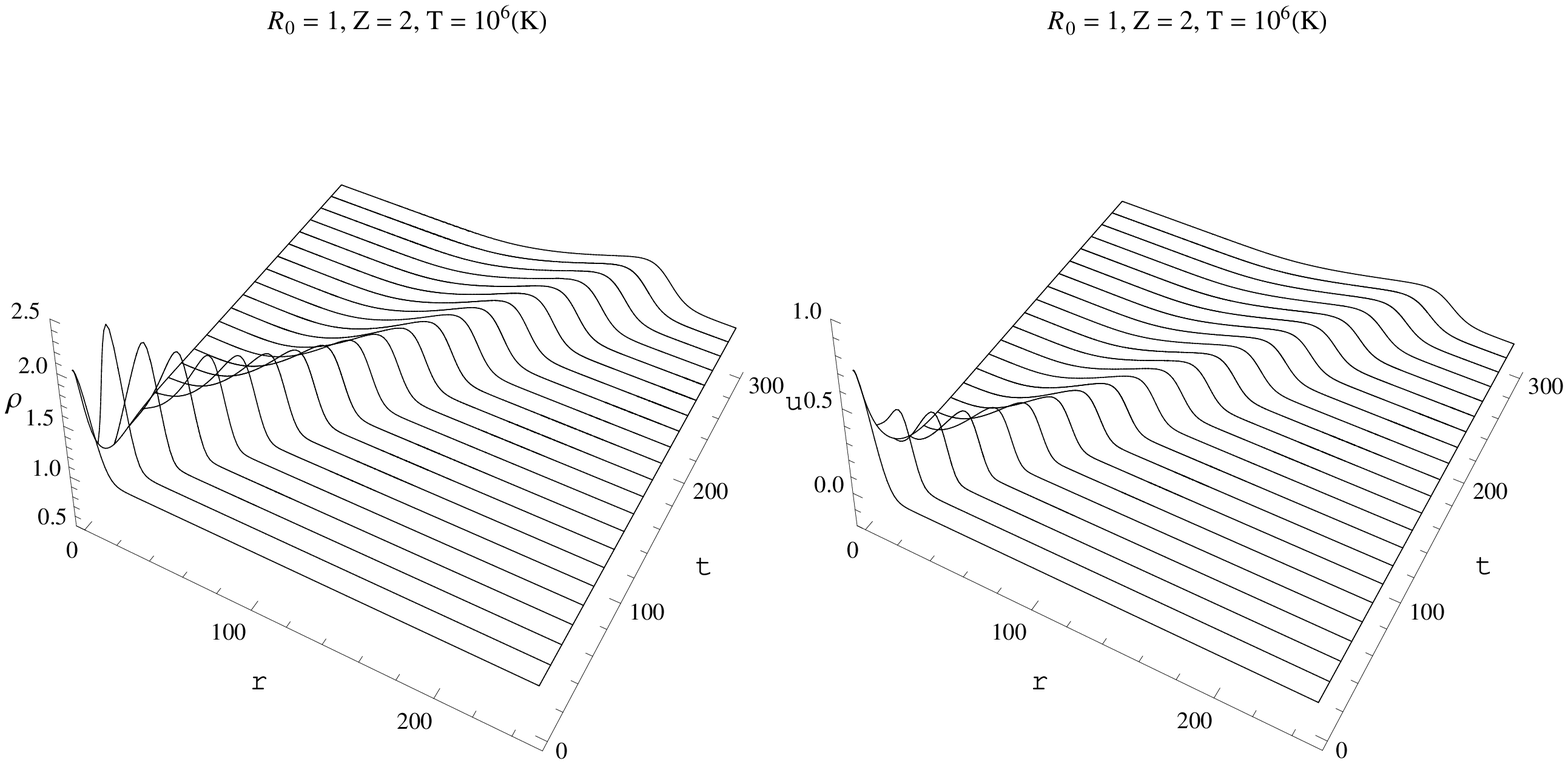}\caption{Figure 4 shows the evolution of a Gaussian density pulse and the corresponding ion-speed profiles in a strongly-coupled collisional quantum plasma (with a given ion mass density, atomic-number and the plasma temperature) into a blast-like nonlinear ion-wave structure.}
\end{figure}

\begin{figure}[ptb]\label{Figure5}
\includegraphics[scale=.6]{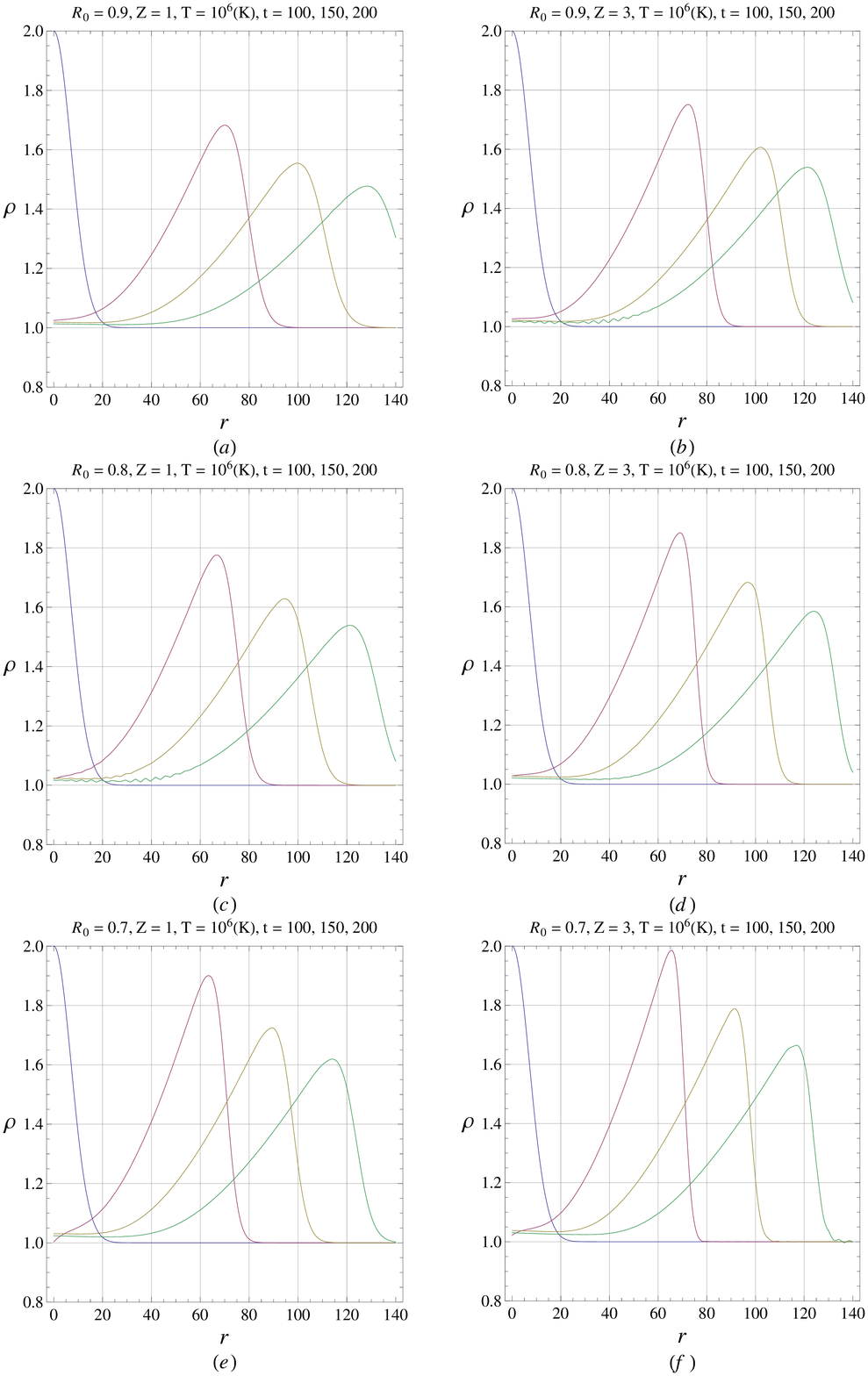}\caption{Figure 5 shows the evolution of a presummed Gaussian density profile in  a strongly-coupled collisional quantum plasma for different simulation times and for different plasma fractional parameters. The figure consists of different plots for variations of the plasma fractional parameters, namely, the ion mass-density (denoted by the relativistic-degeneracy parameter $R_0$) and
the atomic-number $Z$. Each row in the figure indicatess the effect of the variation in atomic number on
the shock height and its steepness and each column depicts the effect of the change in the ion mass density
on the shock wave characteristic parameters.}
\end{figure}

\begin{figure}[ptb]\label{Figure6}
\includegraphics[scale=.6]{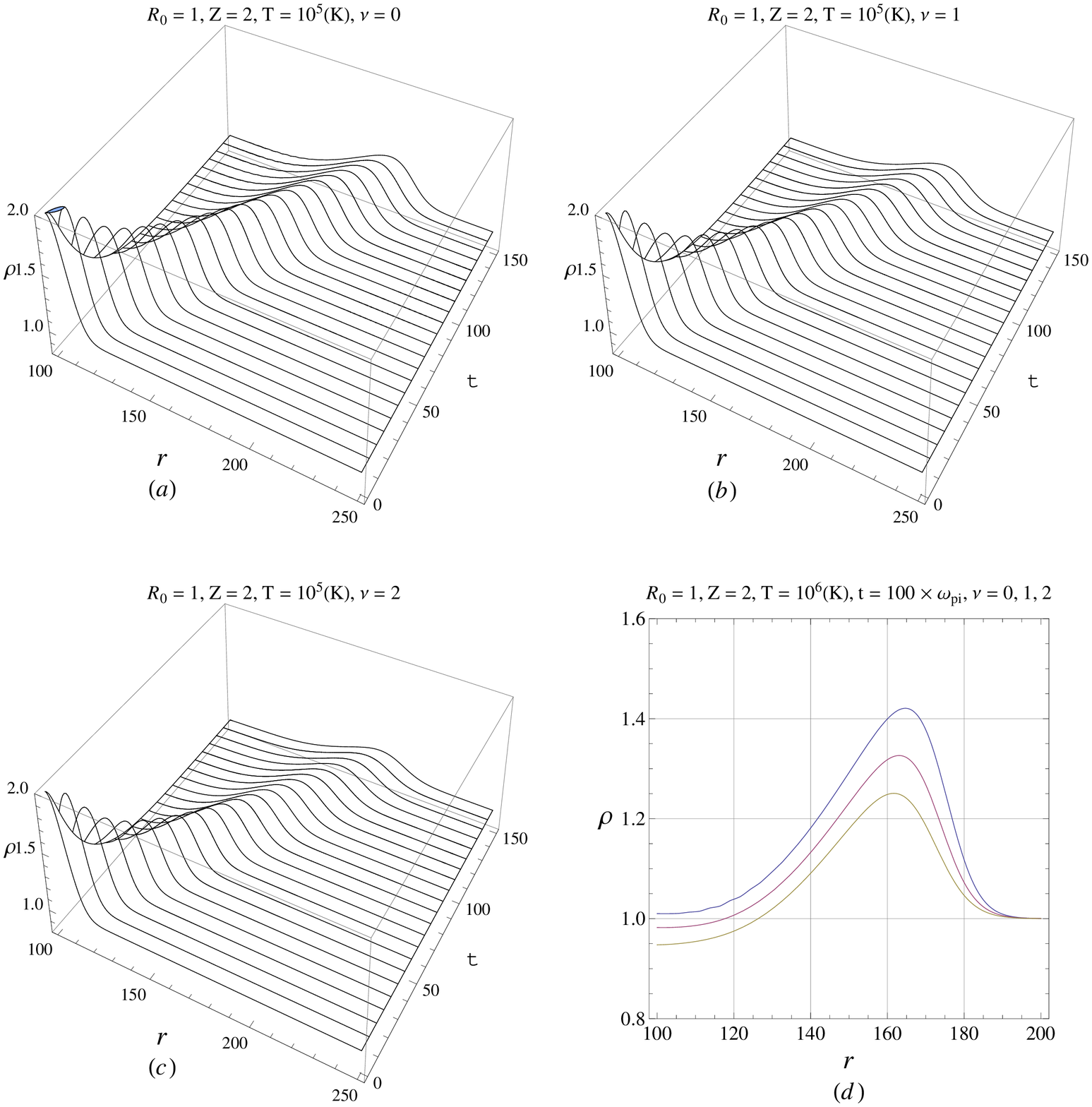}\caption{Figure 6 shows the existence of large-amplitude nonlinear electrostatic shock excitations in different geometries, namely, planar ($\nu=0$), cylindrical ($\nu=1$), and spherical ($\nu=2$). Plot (d) reveals that the shock height increases as one goes from spherical to cylindrical and from cylindrical to planar geometry.}
\end{figure}

It is evident that the generalized set of VEHD equations, Eqs. (\ref{simp}), do not admit analytical arbitrary amplitude solutions. Therefore, at this stage, we aim at numerical investigation of arbitrarily large amplitude nonlinear waves which can have fundamental importance in superdense astrophysical environments. In order to evaluate the nonlinear blast-waves in the laboratory frame, we solve the time-dependent VEHD equations compliant with appropriate periodic boundary conditions for different independent fractional plasma parameter values, namely, the relativistic degeneracy parameter $R_0$ (representing the plasma mass density, $\rho$), and the plasma composition parameter, $Z$. For the the planar geometry (viz. $\nu=0$) a Gaussian pulses of the form ${\rho(r,0)} = 1 + \exp(- {r^2}/100)$ with an initial speed profile of ${u(r,0)} = 0.7\exp(- {r^2}/100)$ is placed at the origin in initial time. The time evolution of the plasma mass-density perturbation is then rendered to future times based on the governing differential equation set, Eq. (\ref{simp}), for given values of the plasma fractional parameters. It is further assumed that the spatial variations in the plasma mass-density vanishes at the plasma boundaries in all future times. In our model, we also incorporate the effects of time-dependence of the viscoelastic parameters due to the plasma mass-density variation, for completeness. The results of simulation are depicted in Figs. 3 and 4 for non-viscous ($\eta=\tau=0$) and viscous ($\eta\neq\tau\neq 0$) quantum plasmas, respectively, for given initial plasma mass-density, temperature and composition relevant to typical white dwarf stars and cores of giant planets. Figure 3 indicates the time evolution of ion density wave into the train of density humps with the time elapse. This is an example of ion wave breaking/steepening, arising from harmonic generation nonlinearity, where, the plasma dispersive effects are somewhat spreading the ion pulse, but not strong enough to allow the formation of a real solitary pulse. These ion density pulses arise from balance between harmonic nonlinearity and the quantum statistical pressure and interaction effects in the limit of vanishing plasma viscosity and the corresponding relaxation time. For the case of strongly coupled quantum plasma the value of the plasma coupling factor, $\Gamma$, is kept below $178$ in our simulations to ensure that the quantum plasma is in the liquid phase \cite{DeWitt}. It is observed from Fig. 4 that the time evolution of the Gaussian density and velocity profiles develop into shock-like structures with pronounced shock wave front steepened by further elapse of  the simulation time.

It is confirmed by Fig. 5 that for all the considered set of plasma parameters the Gaussian perturbation profile develops into shock-like structures after enough elapsed simulation time, say ($t\simeq 100/ \omega_{pi}$). Comparing different plots in Fig. 5, we observe that an increase in both the plasma mass density and the atomic-number leads to an increase in the shock height and steepening of the shock front. Remember that the maximum value of the relativistic degeneracy parameter, $R_0=1$ used in this simulation, corresponds to high electron/ion number density, $n\simeq 5\times 10^{29}cm^{-3}$, expected for white dwarf stars. The present simulation revealed that despite the significant effect of the temperature on the plasma coupling factor, $\Gamma$, it has little effect on the viscosity parameters and the present study might be extended
to the (Jupiter-like) large planet cores and environments with high number densities and relatively low temperatures.

Figure 6 depicts the evolution of ion-momentum distribution from an initial Gaussian profile in different plasma geometries. It is seen that in planar ($\nu=0$), cylindrical ($\nu=1$) and spherical ($\nu=2$) geometries, there exist large amplitude shock waves. In producing Fig. 6(a-c), we have assumed an initial pulse shape as ${\rho(r,0)} = 1 + \exp[- {(r-100)^2}/100]$ and ${u(r,0)} = 0.7\exp[- {(r-100)^2}/100]$ in order
to avoid singularities present at the origin for non-planar geometries. From Fig. 6(d) we observe that the amplitude of nonlinear ion shock structures are relatively higher in the planar geometry with respect to other geometries, for a given set of  fixed plasma parameters. Different geometries than planar, like spherical and cylindrical, may be important in the study of physical mechanisms in stellar configurations and active galactic nuclei, for instance. The present findings reveal the fundamental effects of the plasma mass density and its composition on the formation and characteristics of blast-like ion-waves in dense quantum plasmas. Furthermore, our study reveals that the characteristics of density perturbations and the formation of ES ion shocks in the cores  of white dwarf stars can be quite different from the existing shock-waves at the low density crust region.

\section{Summary and concluding remarks}\label{conclusion}

In this paper, we have presented a nonlinear theory for arbitrary large amplitude electrostatic planar and non-planar shocks in collisional quantum plasmas composed of mildly coupled degenerate electron fluid and strongly correlated non-degenerate ions for a wide range of the plasma mass density and atomic number. We have included the electrostatic and relevant pressure gradient, quantum forces and interactions on degenerate electron fluids, and have used the general VEHD equations for non-degenerate ions to carry out laboratory-frame simulations in different geometries in order to deduce the complete picture of very large-amplitude electrostatic shock structures. The latter may exist in white-dwarfs, the active galactic nuclei or many other high density astrophysical environments such as neutron stars, pulsars and magnetars. It was found that that the increase/decrease in the ion charge state or atomic-number/mass-density would lead to stronger and steeper shock fronts in the relativistically degenerate viscous plasmas. It was also shown that the linear dispersion in such quantum plasmas exhibits anomalous features attributed to the strong fermionic interactions and ion fluid couplings. In conclusion, we stress that the present investigation of the formation of arbitrary large amplitude ES ion shocks will play a very important role in acceleration of electrons and ions in relativistically degenerate quantum plasmas, which are ubiquitous in high-energy density compressed plasmas produced by intense short wavelength laser beams, as well as  in planetary systems  (e.g. the giant Jupiter) and in super dense compact astrophysical objects (e.g. white dwarf stars  and magnetars).

\end{document}